\documentclass[a4paper]{jpconf}
\usepackage{psfrag}
\usepackage{amssymb}
\usepackage{graphicx}
\usepackage{cite}
\begin{document}
\title{Air shower universality from $\mathbf{10^{14}}$ to $\mathbf{10^{22}}$~eV}

\author{A A Lagutin, R I Raikin, T L Serebryakova}

\address{Altai State University, Barnaul, Russia}

\ead{raikin@theory.asu.ru}

\begin{abstract}
Scaling properties of nuclei- and photon-initiated air showers are
examined in wide primary energy range ($10^{14}\div 10^{22}$~eV)
taking into account Landau-Pomeranchuk-Migdal and geomagnetic field
effects. It is shown that the invariance in lateral distribution and
also the universal dependence between shower age and root mean
square radius of electron component exist up to the highest
energies. The implications of shower universality for reanalysis,
cross-checking and validation of the results of different
experiments together with decreasing of the influence of hadronic
model uncertainties are discussed in detail.
\end{abstract}

\section{Introduction}

The concept of air shower universality dating back to pioneering
works~\cite{RossiGreisen,NishimuraKamata,Greisen} is subjected to
the remarkable rise of interest during the last decade~(see
e.g.~\cite{Capdevielle2005,Nerling:2005fj,Giller:2005qz,Gora:2006,Apel:2008hi,Lipari:2008td,Lafebre:2009en,Matthews,Yushkov:2010}).
In a broad sense, the so-called universality is expressed in the
similarity of the spectra of low-energy particles in air showers at
the same development stage. Phenomenology, possible generalisations,
deeper understanding of the origin and limitations of such
properties are obviously important to unveil their potential in the
achievement of the unambiguous physical interpretation of the
results of comparisons between air shower observables and the
appropriate theoretical calculations. Various parameterizations of
the correlations between different air showers characteristics have
been obtained on the basis of universality; attempts have been made
to work out new experimental data processing techniques that could,
for example, suppress or separate the influence of the hadronic
interaction model and improve the reliability of the results.

In this paper we investigate perspectives of model-independent
scaling approach for lateral distribution of electrons in nuclei-
and photon-initiated air showers and also the universal dependence
between shower age and root mean square radius of electron component
for the improved analysis of the experimental data, particulary to
extract information about the variations in elemental composition of
cosmic rays in almost a model-independent way.

\section{Scale-invariance of electron lateral distribution in air showers}

According to the theoretically motivated scaling
approach~\cite{Durban, Predictions}, the electron lateral
distribution function in both electromagnetic cascade showers (EMC)
initiated by high-energy photons and hadron-induced extensive air
showers (EAS) can be described with high accuracy by the
scale-invariant form
\begin{equation}
\rho(r;E,t)=\frac{N(E,t)}{R_0^2(E,t)}\,F\left(\displaystyle\frac{r}{R_0(E,t)}\right).
\label{scaling1}
\end{equation}
Here $\rho(r;E,t)$~-- local particle density at radial distance $r$
from the core in shower of primary energy $E$ observing at depth $t$
in the atmosphere, $N$~-- total number of particles at observation
level (shower size), $R_0$~-- radial scale factor, which, in
contrast to commonly used LDF representations, depends on primary
particle type and mass, shower age and (in case of extensive air
showers) properties of hadronic interactions. Using the results of
extensive series of calculations we found~\cite{Predictions} the
following overall fit for scale-invariant part of lateral
distribution:
\begin{equation}\label{sfnew}
F(x)=Cx^{-\alpha}(1+x)^{-(\beta-\alpha)}(1+(x/10)^\gamma)^{-\delta}.
\end{equation}
For electron LDF in EMC and EAS parameters of function~(\ref{sfnew})
are determined as $C=0.28$, $\alpha=1.2$, $\beta=4.53$,
$\gamma=2.0$, $\delta=0.6$, and scale factor $R_0$ in both cases is
equal to the root mean sqare radius of electron component $R_{\rm
ms}$, defined in a standard way as:
\begin{equation}
R^2_{\rm ms}(E,t)=\frac{2\pi}{N(E,t)}\,
\displaystyle\int_0^\infty r^3 \rho(r;E,t)dr.\label{rms_def}
\end{equation}
The validity of scaling function was confirmed by simulations for
$E=(10^{14}\div 10^{18})$~eV, $t=(600\div 1030)$~g/cm$^2$, $x=(0.05\div 25)$.
The last condition corresponds to the radial distance range from
$r\sim(5\div 10)$~m to $r\sim(2.5\div 4)$~km depending on the shower age.

\section{Scaling LDF and universality property}

It was also found~\cite{Raikin_Germany1,RaikinJPG} that for average
extensive air shower of particular energy, developing in real
atmosphere, there is a one-to-one mapping between $R_{\rm
ms}(E,t)\times\rho(t)$, where $\rho(t)$ is air density at the
observation depth, and $s'=t/(t_{\rm max}+100~{\rm g/cm^2})$, where
$t_{\rm max}$ is the depth of maximum of an average cascade curve.
This allows us to propose the following parameterisation, which does
not depend on energy, observation depth and also primary nuclear
mass:
\begin{equation}
R_{\rm ms}(E,t)=\frac{\rho_0}{\rho(t)}\,A\,
\left[B+\frac{2}{\pi}~{\rm arctg}\left(s'-1\right)\right]~{\rm m},\label{Rms_app}
\end{equation}
where $\rho_0=1.225\times 10^{-3} \rm g/cm^3$, $A=173.0$ and $B=0.546$.

Equation~(\ref{Rms_app}) obviously represents the universal
dependence between average shower width and age. The most remarkable
feature of the described approach is that scaling
property~(\ref{scaling1}), scaling function~(\ref{sfnew}) and
parameterisation~(\ref{Rms_app}) are almost insensitive to hadronic
interactions model implemented in
calculations~\cite{Predictions,Raikin_Germany1,RaikinJPG}.

\begin{figure}[ht]
\begin{minipage}{18pc}
\includegraphics[width=18pc]{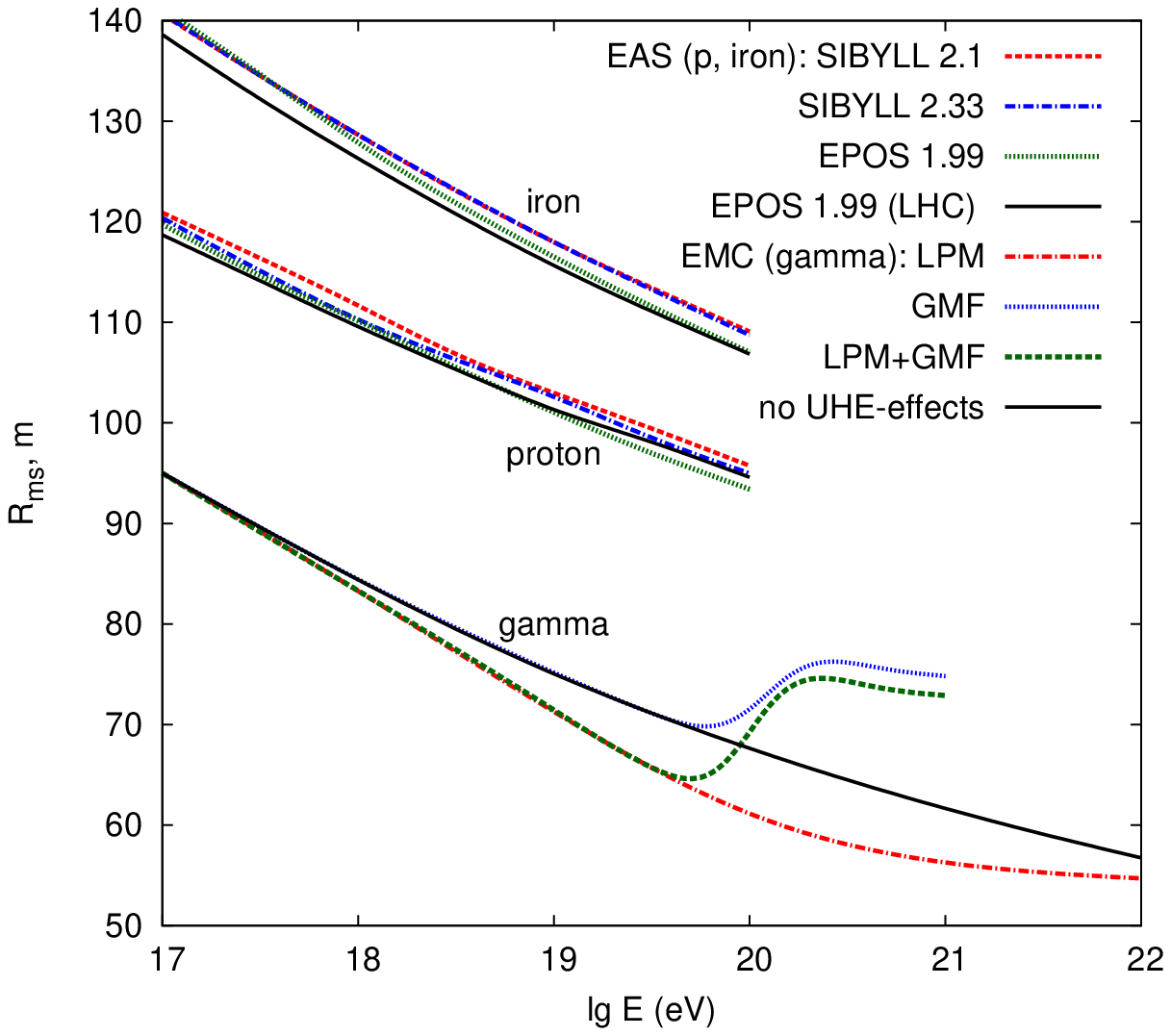}
\caption{\label{rms(e)}Energy dependences of root mean square radius
of electron component in vertical air showers initiated by protons,
iron nuclei (corresponding to recently retuned hadronic models) and
gammas (with and without UHE effects~-- LPM and geomagnetic) at
$890~\rm g/cm^2$ (see text for details).}
\end{minipage}\hspace{2pc}%
\begin{minipage}{18pc}
\includegraphics[width=18pc]{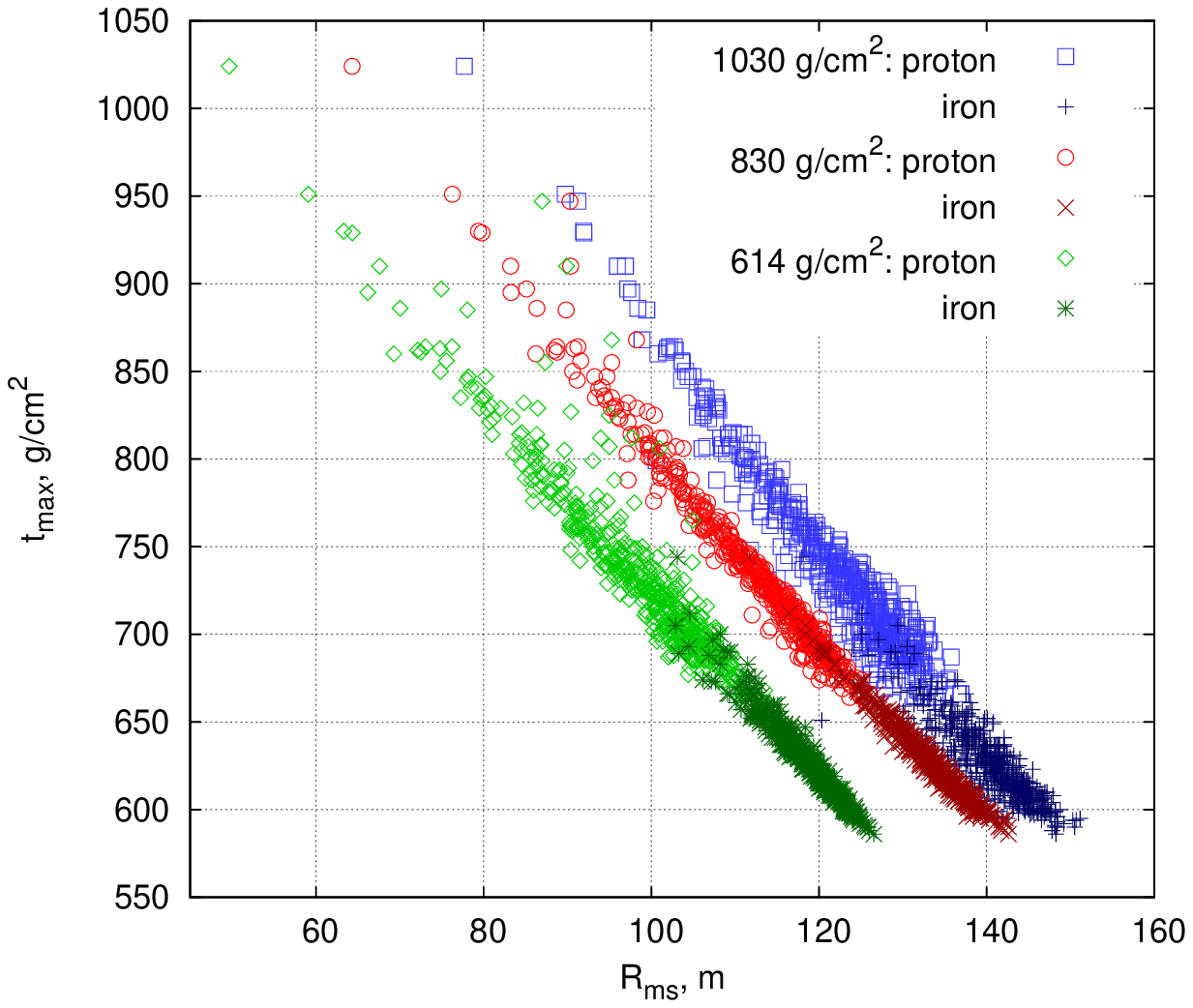}
\caption{\label{XmaxRms}$R_{\rm ms}-t_{\rm max}$ scatter plot for
500 simulated vertical extensive air showers initiated by protons
and iron nuclei with $E=10^{18}$~eV at three observation depths
($614,~830,~1028~\rm g/cm^2$).}
\end{minipage}
\end{figure}

In figure~\ref{rms(e)} energy dependences of root mean square radius
of electron component are shown for vertical EAS initiated by
protons and iron nuclei at observation depth $t=890~\rm g/cm^2$. The
data were obtained from relation~(\ref{Rms_app}) using depths of
maximum calculated with hadronic interactions models retuned on the
basis of recent accelerator data as reported in~\cite{Pierog}. The
uncertainty in $t_{\rm max}$ predictions is currently reduced to
$\sim 20~\rm g/cm^2$ obviously improving the accuracy of conclusions
based on comparisons between simulated and measured depths of
maximum. As it is seen from figure~\ref{rms(e)}, considering radial
scale factors of electron LDF estimated from the experimental data
gives us another important source of information on primary
composition.

We also show in figure~\ref{rms(e)} the mean square radii of pure
electromagnetic cascades initiated by primary gammas examining the
influence of ultra-high energy effects~-- Landau-Pomeranchuk-Migdal
(LPM) effect and interaction of UHE photons and electrons with the
geomagnetic field (GMF). These data were obtained~\cite{Gonch2003,
Gonch2004} on the basis of the numerical solution of adjoint cascade
equations. The GMF intensity profile corresponds to conditions of
Auger Observatory in Argentina. One can see that for UHE primary
gammas $R_{\rm ms}$, keeping much smaller values comparing to EAS,
increases considerably with energy from $E\sim 3\times 10^{19}~\rm
eV$ to $E\sim 3\times 10^{20}~\rm eV$ after interaction with
geomagnetic field begins to affect shower development. This should
be taken into account when $R_{\rm ms}$ is used for testing
gamma-ray fraction in UHE cosmic rays flux.

On the other hand, our calculations show that scaling
formalism~(\ref{scaling1})-(\ref{sfnew}) remains valid when LPM- and
GMF-effects in EMC are taken into account~\cite{Gonch2003,
Gonch2004}. Since for electron LDF in UHE extensive air showers only
LPM-effect in partial electromagnetic cascades contributes to
$R_{\rm ms}$ we expect that, barring dramatic changes in nuclear
interactions, both scaling formalism~(\ref{scaling1})-(\ref{sfnew})
and universality in form~(\ref{Rms_app}) will be kept up to
$10^{22}$~eV.

\section{Implications for experimental data analysis at different energies}

In our recent works~\cite{RaikinASU2010, Raikin:2008zz,
Raikin:2009zz, Raikin:2011zz} the experimental data of KASCADE,
KASCADE-Grande and MSU EAS arrays were analyzed with respect to the
radial scale factors of electron lateral distribution in the
framework of generalized scaling formalism taking into account the
appropriate shower classification procedure.

It was shown that uncertainty in the explicit form of lateral
distribution function chosen for data fitting and bias of different
nature in estimating the mean square radius of the EAS electron
component do not impede the extraction of information about mean
primary mass change with energy. The variations of $R_0$ with energy
obtained  using different model LDFs from the experimental data of
the above mentioned arrays give a consistent evidence of weighting
of primary composition above the knee. According to the
KASCADE-Grande data it is followed by a rather sharp decrease of
mean primary mass at $\lg N_e\gtrsim 7.5$~\cite{Raikin:2011zz},
indicative in favor of a ``heavy knee" in primary spectrum.

Considering the applicability of such an analysis at very high
energies for the experimental lateral distributions measured by
largest air shower arrays it is important in what extent the
universality property is affected by fluctuations in individual
showers. In figure~\ref{XmaxRms} the $R_{\rm ms}-t_{\rm max}$
scatter plot is presented for 500 simulated vertical extensive air
showers initiated by protons and iron nuclei with $E=10^{18}$~eV at
three observation depths ($614,~830,~1030~\rm g/cm^2$). Observation
depths are chosen in such a way as to analyze  $R_{\rm ms}-t_{\rm
max}$ correlations in individual showers of different age. It is
clear that for the majority of individual EAS of particular energy
$614~\rm g/cm^2$ level is located higher than depth of maximum,
while sea level is considerably deeper than $t_{\rm max}$. At the
same time $830~\rm g/cm^2$ roughly corresponds to the depth where
fluctuations in shower size are minimal. Dependence on the
observation level could also be used to infer the dependence on the
shower zenith angle.

It is seen from figure~\ref{XmaxRms} that strong correlation between
root mean square radius and depth of maximum in individual extensive
air showers is observed for showers of different ages and for both
light and heavy primary nuclei, which, in addition, are separated
quite well from each other on scatter plot. This fact suggests that
implementation of radial scale factors into multicomponent analysis
procedures, e.g. the principal component analysis, could be more
efficient comparing with utilizing of other observables sensitive to
primary particle type such as electron and muon shower sizes and
local densities far from shower core position. Of course, additional
studies devoted to the reliable estimation of radial scale factors
of electron LDF from the experimental data of largest ground-based
air shower arrays taking into account realistic measurement
uncertainties, detectors response and other experimental conditions
are needed. This is essential especially for hybrid arrays measuring
longitudinal and lateral profiles of the shower independently.

\section{Conclusion}
The universality in extensive air shower development expressed by
means of scale-invariant lateral distribution of electrons and the
relation between root mean square radius and depth of shower maximum
is evidently retaining in a common form in extremely wide primary
energy range (from $10^{14}$ to $10^{22}$~eV) neglecting possible
exotic hadronic phenomena. This formalism could be useful in
development of new effective techniques for reanalysis,
cross-checking and validation of the results of different
experiments together with decreasing of the influence of hadronic
model uncertainties allowing to raise confidence in the results
coming from the comparisons between air showers observables and
theoretical calculations.


\section*{References}

\begin{thebibliography}{99}

\bibitem{RossiGreisen}Rossi B, Greisen K 1941 {\it Rev. Mod. Phys.} {\bf 13} 240

\bibitem{NishimuraKamata}Nishimura J, Kamata K 1958 {\it Progr. Theor. Phys.} {\bf 6} 93

\bibitem{Greisen}Greisen K 1960 {\it Ann. Rev. Nucl. Part. Sci.} {\bf 10} 63

\bibitem{Capdevielle2005}Capdevielle J N. and Cohen F 2005 {\it J.\ Phys.\ G} {\bf 31} 507

\bibitem{Nerling:2005fj}Nerling F et al 2006 {\it Astropart. Phys.} {\bf 24} 421

\bibitem{Giller:2005qz}Giller M et al 2005 {\it J.\ Phys.\ G} {\bf 31} 947

\bibitem{Gora:2006}Gora D et al 2006 {\it Astropart. Phys} {\bf 24} 484

\bibitem{Apel:2008hi}Apel W D et al 2008 {\it Astropart. Phys.} {\bf 29} 412

\bibitem{Lipari:2008td}Lipari P 2009 {\it Phys.\ Rev.\  D} {\bf 79} 063001

\bibitem{Lafebre:2009en}Lafebre S et al 2009 {\it Astropart.\ Phys.} {\bf 31} 243

\bibitem{Matthews}Matthews J A J et al 2010 {\it J. Phys. G: Nucl. Part. Phys.} {\bf 37} 025202

\bibitem{Yushkov:2010}Yushkov A et al 2010 {\it Phys. Rev. D} {\bf 81} 123004


\bibitem{Durban}Lagutin A A et al 1997 {\it Proc. 25 ICRC (Durban)} {\bf V.\,6} 285-288

\bibitem{Predictions}Lagutin A A, Raikin R I 2001 {\it Nucl. Phys. B (Proc. Suppl.)} {\bf 97B} 274-277


\bibitem{Raikin_Germany1}Raikin R I et al 2001 {\it Proc. 27 ICRC (Hamburg)} {\bf V.\,1} 290-293, 294-297

\bibitem{RaikinJPG}Lagutin A A et al 2002 {\it J. Phys. G: Nucl. Part. Phys.} {\bf 28} 1259-1274

%%\bibitem{Pryke}Pryke C 2001 {\it Astropart. Phys.} {\bf 14} 319. astro-ph/0003442

\bibitem{Pierog}Pierog T 2012 {\it Proc. Int. Symp. on Future Directions in UHECR Physics (CERN)}

\bibitem{Gonch2003}Goncharov A I et al 2003 {\it Proc. 28 ICRC (Tsukuba)} {\bf V.\,1} 575-578

\bibitem{Gonch2004}Goncharov A\,I et al. 2004 {\it Russian Physics Journal} {\bf 47} N4 444-452

\bibitem{RaikinASU2010}Raikin R I, Lagutin A A 2008 {\it Izv. Altai Gos. Univ.}  {\bf 1} 66-71

\bibitem{Raikin:2008zz}Raikin R I et al 2008 {\it Nucl.\ Phys. B (Proc.\ Suppl.)} {\bf 175-176} 559-563

\bibitem{Raikin:2009zz}Raikin R I et al 2009 {\it Nucl.\ Phys. B (Proc.\ Suppl.)} {\bf 196} 383-386

\bibitem{Raikin:2011zz}Raikin R I, Lagutin A A 2011 {\it Proc. 32 ICRC (Beijing)} {\bf V.\,1} 299-302

\end{thebibliography}
\end{document}